\documentclass[aps,prd,reprint,superscriptaddress, showpacs]{revtex4-2}

\usepackage{graphicx}
\usepackage{amsmath}
\usepackage{color}

\begin{document}
\title{\textbf{Proposal for the search for exotic spin-spin interactions at the micrometer scale using functionalized cantilever force sensors}}

\author{Qian Wang}
\affiliation{MOE Key Laboratory of Fundamental Quantities Measurement, Hubei Key Laboratory of Gravitation and Quantum Physics, Huazhong University of Science and Technology, Wuhan 430074, China}

\author{Ze Ouyang}
\affiliation{School of Physics, Huazhong University of Science and Technology, Wuhan 430074, China}

\author{Yu Lu}
\affiliation{MOE Key Laboratory of Fundamental Quantities Measurement, Hubei Key Laboratory of Gravitation and Quantum Physics, Huazhong University of Science and Technology, Wuhan 430074, China}

\author{Jianbo Wang}
\affiliation{Huazhong Institute of Electron-Optics, Wuhan National Laboratory for Optoelectronics, Wuhan 430223, China}

\author{Lin Zhu}
\affiliation{MOE Key Laboratory of Fundamental Quantities Measurement, Hubei Key Laboratory of Gravitation and Quantum Physics, Huazhong University of Science and Technology, Wuhan 430074, China}

\author{Pengshun Luo}
\email{pluo2009@hust.edu.cn}
\affiliation{MOE Key Laboratory of Fundamental Quantities Measurement, Hubei Key Laboratory of Gravitation and Quantum Physics, Huazhong University of Science and Technology, Wuhan 430074, China}

\date{\today}

\begin{abstract}

Spin-dependent exotic interactions can be generated by exchanging hypothetical bosons, which were introduced to solve some puzzles in physics. Many precision experiments have been performed to search for such interactions, but no confirmed observation has been made. Here, we propose new experiments to search for the exotic spin-spin interactions that can be mediated by axions or \textit{Z$^\prime$} bosons. A sensitive functionalized cantilever is utilized as a force sensor to measure the interactions between the spin-polarized electrons in a periodic magnetic source structure and a closed-loop magnetic structure integrated on the cantilever. The source is set to oscillate during data acquisition to modulate the exotic force signal to high harmonics of the oscillating frequency. This helps to suppress the spurious signals at the signal frequency. Different magnetic source structures are designed for different interaction detections. A magnetic stripe structure is designed for \textit{Z$^\prime$}-mediated interaction, which is insensitive to the detection of axion-mediated interaction. This allows us to measure the coupling constant of both if we assume both exist. With the force sensitivity achievable at low temperature, the proposed experiments are expected to search for the parameter spaces with much smaller coupling constant than the current stringent constraints from micrometer to millimeter range. Specifically, the lower bound of the parameter space will be seven orders of magnitude lower than the stringent constraints for \textit{Z$^\prime$}-mediated interaction,  and an order of magnitude lower for axion-mediated interaction, at the interaction range of $10\, \mu$m.

\end{abstract}

\maketitle

\section{Introduction}

The searches for spin-dependent exotic interactions have recently attracted attention in particle physics related fields\cite{Adelberger2009, Safronova2018, Ficek2022}. These interactions can occur between two fermions by exchanging new spin-0 or spin-1 bosons\cite{Moody1984, Fayet1986, Fayet1996, Dobrescu2005, Dobrescu2006, Fadeev2019}, which have been proposed to address some mysteries in physics, such as the strong CP problem \cite{Peccei1977, Weinberg1978, Wilczek1978, Kim2010}, dark matter \cite{Bertone2005, Arkani2009}, dark energy \cite{Copeland2006, Peebles2003, Kamionkowski2014},  and hierarchy problem \cite{Arkani1998, Graham2015}. Among them, the axion is one of the well-motivated bosons introduced to solve the strong CP problem and is now a promising candidate for dark matter\cite{DiLuzio2020, Graham2015axion, Sikivie2021}. As Moody and Wilczek first pointed out, spin-dependent exotic interactions can arise through axion exchange\cite{Moody1984}. In a more general discussion by Dobrescu and Mocioiu, the spin-dependent potentials were classified into 15 types by their mathematical spin-momentum structures\cite{Dobrescu2006}.  These potentials have recently been re-derived  in a form that clearly shows  the relationship between the potentials and the bosons mediating them\cite{ Fadeev2019}, and shows that  the interactions can be generated by pseudoscalar coupling, vector coupling, and axial vector coupling between fermions and generic spin-0 or spin-1 bosons. 

In this paper, we propose new experiments to explore the following spin-spin interactions between electrons, enumerated $V_2$ and $V_3$ in Ref. \cite{Dobrescu2006},
\begin{equation}
    V_2=\frac{g_A^{e}g_A^{e}}{4\pi \hbar c}\frac{\hbar c}{r}\left(\hat \sigma_1 \cdot \hat \sigma_2\right)e^{-r/\lambda}, \label{v2}
\end{equation}
\begin{equation}
\begin{aligned}
    V_3=&-\frac{g_p^{e}g_p^{e}}{4\pi \hbar c}\frac{\hbar^3}{4m_e^2c}\left[\left(\hat \sigma_1 \cdot \hat \sigma_2\right) \left(\frac{1}{\lambda r^2}+\frac{1}{r^3} \right)-\right.\\ &\left. \left(\hat \sigma_1 \cdot \hat r \right)\left(\hat \sigma_2 \cdot \hat r \right)\left(\frac{1}{\lambda^2 r}+\frac{3}{\lambda r^2}+\frac{3}{r^3}\right) \right] e^{-r/\lambda}, \label{v3}
\end{aligned}
\end{equation}
where ${g_A^{e}g_A^{e}}/{4\pi \hbar c}$ and  ${g_p^{e}g_p^{e}}/{4\pi \hbar c}$ are the dimensionless coupling constants, $\hbar$ is the Dirac constant, $c$ is the speed of light in vacuum, $\hat \sigma_1$ and $\hat \sigma_2$ are the unit spin vectors of the electrons, $r$ is the distance between them, $\hat r$ is the unit relative position vector, and $\lambda$ is the interaction range. Here $\lambda=\hbar/m_b c$ is the reduced Compton wavelength of the hypothetical boson that mediates the interaction, and $m_b$ is its mass. The $V_2$ potential can be mediated by a spin-1 \textit{Z$^\prime$} boson via axial-vector coupling\cite{Fayet1986, Fayet1996, Dobrescu2006, Fadeev2019}. The $V_3$ potential can be mediated by spin-0 pseudoscalar bosons, such as axions or axion-like particles\cite{Moody1984, Dobrescu2006, Fadeev2019}.

Various techniques have been applied or proposed to search for these exotic potentials, including atomic and optical precision measurement\cite{Wineland1991, Glenday2008, Vasilakis2009, Ledbetter2013, Hunter2013, Kotler2015, Luo2017, Ji2017, Ficek2017, Ficek2018, Rong2018, Almasi2020, Wang2022}, mechanical sensors\cite{Ritter1990, Heckel2013, Hunter2013, Terrano2015, Leslie2014}, and SQUIDs\cite{Chui1993}. So far, there has been no convincing evidence for the existence of new interactions, but experiments have placed increasingly stringent constraints on them. For the $V_2$ interaction in the interaction range from 0.1 $\mu$m to 1 mm, the most stringent constraints are set by the experiments with trapping strontium ions\cite{Kotler2015} and quantum diamond sensors\cite{Rong2018}. The analysis of helium atomic spectra has been used to impose the strictest constraints on $V_3$ interaction\cite{Ficek2017}. The above constraints have been obtained by comparing the experimental data with the theoretical calculation of magnetic dipole-dipole interaction. The results depend on the experimental measurement noise, the accuracy of the theoretical calculation, and how well the experimental data matches the theoretical values.

Here we propose to search for the exotic interactions by measuring the force between two magnetized objects with a cantilever. To avoid the high precision requirement for calculating the electromagnetic effects, we employ periodic magnetic structures that can generate spatially varied exotic force signals, so that we can distinguish the signals of interest from interfering forces. For another interacting object, a closed magnetic loop enclosed with superconducting thin film shielding is used to suppress the magnetic force. Different periodic magnetic structures are designed for different interaction detections, which enables us to perform joint data analysis under the assumption that both $V_2$ and $V_3$ could exist, whereas each was usually considered independently in previous literature. Finally, using a sensitive cantilever allows us to probe the exotic interactions at distances 
in the range of micrometers with high precision.

This paper is organized as follows. Sec. \ref{experimental scheme} illustrates the experimental scheme. Sec. \ref{experimental design} describes the experimental designs, including the probe and source structures in details, as well as the expected force signal and parameter space that can be explored. In Sec. \ref{error analysis}, we discuss the influence of the spurious forces likely to appear in the experiments. The conclusions are given in Sec. \ref{conclusion}.

\section{Experimental scheme}\label{experimental scheme}

\begin{figure}[htbp]
    \centering
    \includegraphics[width=8.5cm]{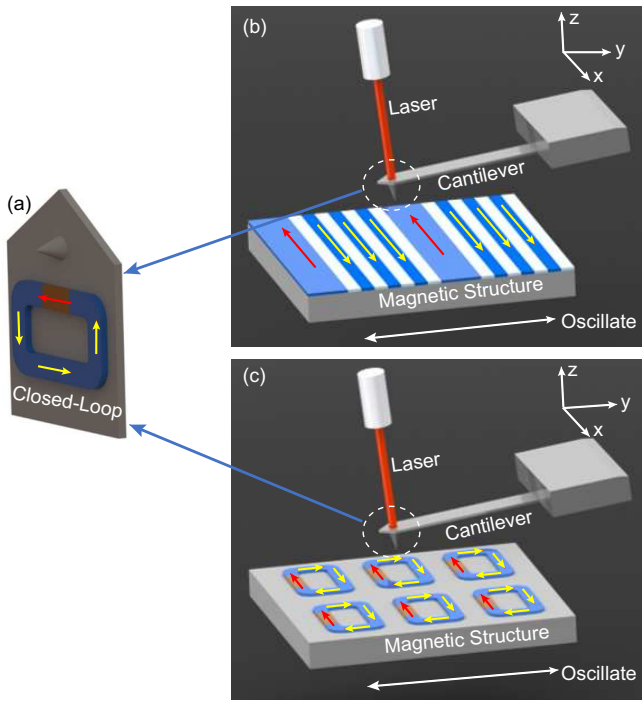}
    \caption{Schematic of the proposed experiments. (a) The end part of the cantilever with the CLMS integrated. The arrows indicate the direction of magnetization, where the yellow arrows represent the magnetization of the soft magnet, and the red arrow represents the magnetization of the permanent magnet. (b) The proposed experiment searches for $V_2$ interaction. A fiber interferometer is used to measure the displacement of the cantilever. The spin-polarized source is designed as alternative antiparallel spin-polarized magnetic stripes. (c) The proposed experiment searches for $V_3$ interaction. The spin-polarized source is designed as a periodic array of the closed-loop magnetic structures.}
    \label{fig:exp_scheme}
\end{figure}

The experiments are schematically shown in Fig. \ref{fig:exp_scheme}. A cantilever is used as a force sensor to measure the exotic interaction between the spin-polarized electrons in the closed-loop magnetic structure (CLMS) on the cantilever and that in another source separated by several micrometers from each other. The source is a periodic magnetic structure, which is expected to produce a spatially periodic exotic potential field. Thus once the source is driven to oscillate by a piezo element, a time-varying force is expected to exert on the cantilever and make it oscillate. The displacement of the cantilever can be measured by a fiber interferometer. In the frequency domain, the mechanical response of a force acting on the cantilever is
\begin{equation}
    z(\omega)=\frac{1}{m}\frac{F_z(\omega)}{\omega_0^2-\omega^2+\frac{i\omega \omega_0}{Q}}\label{x(omega)},
\end{equation}
where the subscript $z$ indicates the force along the $z$-axis, $z(\omega)$ denotes the displacement of the cantilever in the frequency domain, $\omega_0$ is the intrinsic resonant angular frequency of the cantilever, $Q$ is the quality factor of the cantilever, and $m$ denotes the total effective mass of the cantilever. 

The exotic force $F_z$ is calculated by
\begin{equation}
    F_z=-\frac{\partial}{\partial d}\int n_sn_pV(r)\mathrm{d}V_s\mathrm{d}V_p \label{F_z},
\end{equation}
where $n_s$ is the number density of the spin-polarized electrons in the periodic source structure, and $n_p$ is that in the CLMS. The integral is performed on the exotic potential $V(r)$ over both volumes of the source ($V_s$) and CLMS ($V_p$). The force is obtained by taking the derivative of the integral with respect to $d$, the distance between the CLMS and the spin-polarized source. 

The sources are specially designed for different exotic interactions. We use magnetic stripes with periodic antiparallel spin-polarization to detect the exotic potential $V_2$ [see Fig.\ref{fig:exp_scheme} (b)]. This structure can generate a periodic $V_2$ signal, while creating negligible $V_3$ force if we make the stripes sufficiently long. The magnetic field generated by the magnetic stripes lies in the plane and closes at the end of the stripes, so that the magnetic field produced at the CLMS is small and the induced magnetic force is negligible. However, the $V_2$ potential decays exponentially with distance so that only the segments of the stripes near the CLMS contributed to the force.  Another structure, made of CLMS array, is used for the detection of $V_3$ [see Fig.\ref{fig:exp_scheme} (c)]. It should be noted that this structure also generates $V_2$ signal, so that we can combine the two experiments to measure the strength of both interactions assuming the presence of both. To reduce the disturbance of the Casimir force and electrostatic force, the surfaces of the sources are coated with a layer of metallic thin film or superconducting thin film.

\section{Experimental Design}\label{experimental design}

\subsection{Cantilever with a closed-loop magnetic structure}\label{probe}
\begin{figure}[htbp]
    \centering
    \includegraphics[width=7.3cm]{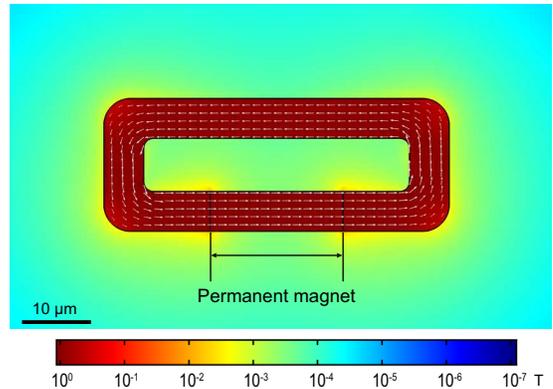}
    \caption{FEA simulation of the magnetic properties of the CLMS. The arrows indicate the magnetization in the central plane of the CLMS, and the color scale shows the magnitude of the magnetic flux density. A magnetization of 800 kA/m is used for the permanent magnet, the soft magnet is simulated with a relative magnetic susceptibility of 8000.}
    \label{fig:mag_loop}
\end{figure}

Searching for the spin-spin interactions requires the use of spin-polarized objects, thus the magnetic force between the objects is a key factor to consider. To reduce the stray field produced by the object, we consider using a cantilever with a CLMS attached at its end. The CLMS is made of a soft magnetic loop (e.g. $\mathrm{Ni_{80}Fe_{20}}$) with a permanent magnetic segment (e.g. $\mathrm{SmCo_{5}}$) embedded in it, as shown in Fig. \ref{fig:exp_scheme} (a) and Fig. \ref{fig:mag_loop}. The permanent magnet can magnetize the soft magnetic material, and the electron spins are then polarized along the loop, providing the source of electron spins for the spin-spin interactions. As the magnetization is roughly closed in a loop, the CLMS creates a tiny stray field outside it. 

The finite element analysis (FEA) is conducted to simulate the magnetization and stray field of the CLMS. Figure. \ref{fig:mag_loop} shows the simulated distribution of the magnetic flux density at its remnant state. We can see that a toroidal magnetization forms, except for a relatively small  leakage magnetic field around the junctions between the two different materials. The leakage magnetic field is on the order of mT, which can create a magnetic force larger than the force sensitivity of the cantilever in the $V_3$ search experiment. Since the leakage magnetic field is smaller than the lower critical field of the NbTi superconductor, it can be shielded by enclosing the CLMS inside the NbTi thin films. According to the simulation, using 1.5-$\mu m$-thick NbTi thin film can shield the magnetic field down to $10^{-8}$ T, which will be discussed in details in Sec. \ref{error analysis}. 

The magnetic loops can be micro-fabricated on a silicon on insulator (SOI) wafer with NbTi thin film pre-deposited. After the magnetic loops are fabricated, another NbTi layer is deposited on the structure to enclose all the magnetic materials. By selectively etching off the handle layer of the SOI wafer, we can leave the CLMS on the suspended silicon device layer, which enables us to cut the structure with focused ion beam (FIB), and then transfer it to a customized cantilever with a tip height of $\sim$ 10 $\mu$m.

\subsection{Minimum detectable force}
The minimum detectable force depends on the thermal noise of the cantilever and the displacement measurement noise of the fiber interferometer. The thermal noise of the cantilever is given by 
\begin{equation}
    S_{F_T}^{1/2}\left(f\right)=\sqrt{\frac{2kk_BT}{\pi fQ}}, \label{thermal noise}
\end{equation}
where $k$ is the spring constant of the cantilever chosen to be $0.02\,\mathrm{N/m}$, $k_B$ is the Boltzmann constant, $T$ is the temperature, and $Q$ is the quality factor of the cantilever. The experiments need to be conducted at low temperature for superconducting shielding to work.  Using the base temperature (6 K) of our instrument, we calculate the thermal noise to be 2.0$\times10^{-15}\mathrm{N/\sqrt{\mathrm{Hz}}}$ by conservatively assuming $Q= 10000$. The displacement measurement noise of  100 $\mathrm{fm/\sqrt{\mathrm{Hz}}}$ can be achieved at the frequency of interest. Given the acquisition time of 1000 $s$ and signal frequency of 25.8 Hz\cite{Ding2020}, the minimum detectable force is estimated to be 8.9$\times10^{-17}$   $\mathrm{N}$ as the quadrature sum of the two contributions.

\subsection{Search for $V_2$ interaction}\label{New limit on v2}

To search for the $V_2$ interaction, we use periodic magnetic stripes of different widths as another source [see Fig. \ref{fig:exp_scheme}(b)]. Since the coercive field of the narrow stripes is larger than that of the wide stripes due to shape dependent demagnetization, the magnetic structure can be prepared in an antiparallel state in the following way. First let us apply a magnetic field large enough to magnetize all the stripes in the same direction, say $+x$ direction, then we reverse the field to just flip the magnetization of the wide stripes. Since each stripe has a near square hysteresis loop, the antiparallel state remains after removing the magnetic field. Such structures were successfully fabricated in the previous experiment \cite{Ding2020}, where their surfaces  are further coated with gold films to reduce the contribution of the Casimir force and electrostatic force. 

\begin{figure}[htbp]
   \centering
   \includegraphics[width=8.5cm]{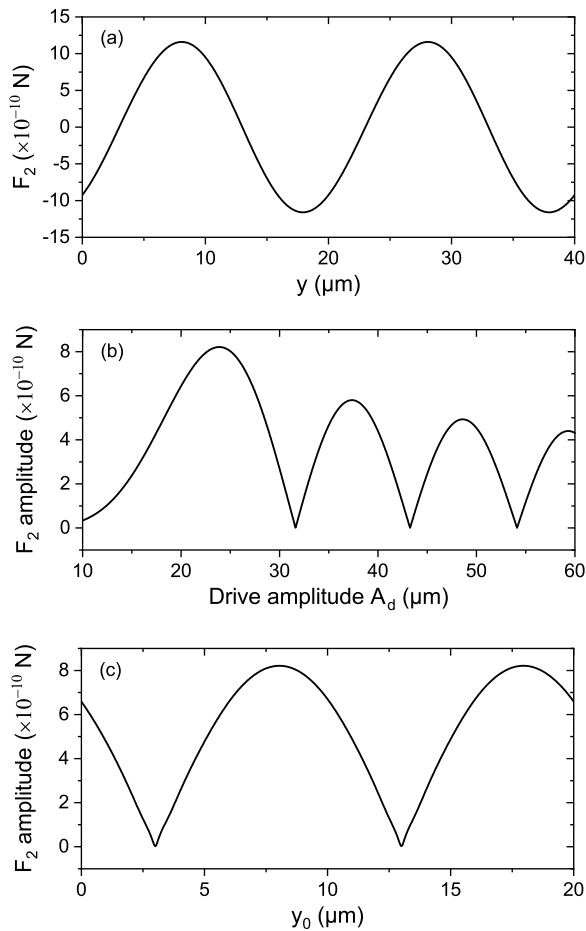}
  \caption{(a) The expected $V_2$ force varies with the relative position along the $y$-direction. (b) The $V_2$ force amplitude at $6f_d$ as a function of the driving amplitude.  (c) The $V_2$ force amplitude at $6f_d$ as a function of $y_0$. In the calculation, $g_A^eg_A^e/4\pi \hbar c$ is set to $1.8\times 10^{-19}$ with $\lambda$ = 10 $\mu$m.}
    \label{fig:v2_force}
\end{figure}

The preliminary design parameters of the structure are listed in Table \ref{tab:parameter}. The expected $V_2$ force is numerically calculated as a function of the lateral position $y$ for $\lambda = 10\,\mu$m, the result is shown in Fig. \ref{fig:v2_force} (a). Here the coupling constant $g_A^eg_A^e/4\pi \hbar c$ is chosen to be $1.8\times 10^{-19}$, which is the most stringent constraint given by the experiment based on quantum diamond sensors so far. The number density of spin-polarized electrons $n$ in the structure is given by
\begin{equation}                                         n=\frac{Mr_{sa}}{\mu_B} \label{the density of spin electron number},
\end{equation}
where $M$ is the magnetization of the CLMS, $\mu_B$ is the Bohr magneton, and $r_{sa}$ is the ratio of the spin to all magnetic moments, depending on the material composition \cite{Glaubitz2011}. The $V_2$ force is periodic with $y$, and varies with an amplitude of $8.2\times 10^{-10}$ N, which is about 7 orders of magnitude larger than the minimum detectable force of the cantilever. During data acquisition, we drive the source to oscillate as $y=y_0+A_d\mathrm{cos}\left(2\pi f_dt\right)$ and record the resulting time-varying signal. The exotic force signal is then modulated to the harmonic frequencies, which helps us separate the spurious signals from the signal of interest. The exotic force amplitude at the m$th$ harmonic frequency is given by
\begin{equation}
\begin{aligned}
    F_{m}(y_0) =\sum_{n=-\infty}^{+\infty} 
    i^{m}J_{m}(k_nA_d)\left[f(k_n)e^{ik_ny_0} \right],
    \label{harmonics} 
\end{aligned}
\end{equation}
where $f(k_n)$ is the n$th$ coefficient of the Fourier series expansion of $F_z\left(y\right)$, $k_n=n2\pi/\Lambda$ and $\Lambda$ is the magnetic structure period, $J_{m}$ is the Bessel function of order $m$. Fig. \ref{fig:v2_force} (b) shows the $V_2$ force amplitude at $6f_d$ as a function of driving amplitude $A_d$. We can see that the optimal value for $A_d$ is $23.9\,\mu$m, which maximizes the force amplitude at $6f_d$. 

\begin{figure}[htbp]
    \centering
    \includegraphics[width=8.0cm]{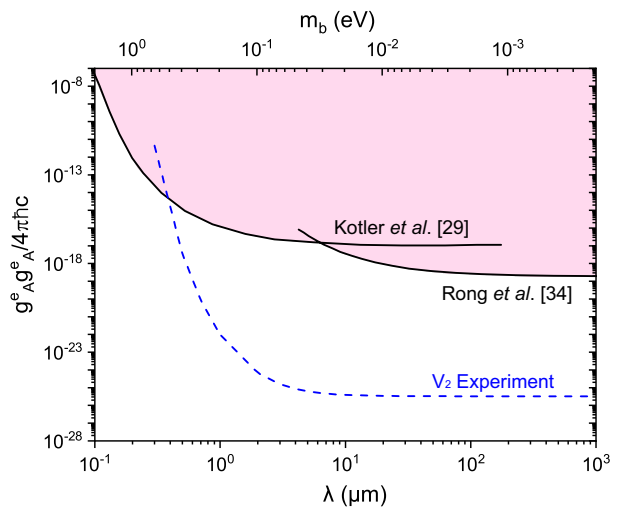}
    \caption{The constraints on the coupling constant of the $V_2$ potential. The dashed line represents the lower bound of the parameter space that the proposed experiment can explore.}
    \label{fig:v2_coupling}
\end{figure}

The force amplitude is a periodic function of $y_0$, the equilibrium position of the oscillation. Therefore, we can collect data by changing $y_0$ over a range larger than one period, and the expected result is shown in Fig. \ref{fig:v2_force} (c). If we do not observe any periodic signal in such measurement, the $V_2$ force must be lower than the minimum detectable force. Based on the preliminary design parameters, the potential limit on the coupling constant $g_A^eg_A^e/4\pi \hbar c$ can be obtained, which is shown in Fig. \ref{fig:v2_coupling}. The result indicates that we can explore a range of coupling constant down to 7 orders of magnitude lower than the current strictest constraint at $\lambda=10\,\mu$m.

\subsection{Search for $V_3$ interaction}
The magnetic stripe structure is not a suitable source for the search for the $V_3$ interaction. The $V_3$ force between the CLMS and the stripe structure is greatly suppressed because of the subtracting terms in Eq. (\ref{v3}) canceling each other for sufficient long magnetic stripes. That makes the magnetic stripe structure only sensitive to $V_2$ detection. To search for the $V_3$ interaction, we need to cut the stripes into segments with optimal length and spacing. To keep the magnetic force low, we choose to use the CLMS array as the source for $V_3$ detection, as shown in Fig. \ref{fig:exp_scheme} (c). Each CLMS in the array has the same dimensions as the CLMS on the cantilever, the spacing between them is optimized, and the values are listed in Table \ref{tab:parameter}. To further reduce the magnetic force below the minimum detectable force, the CLMS array needs to be shielded by superconducting films, which will be discussed in Sec. \ref{spurious, magnetic force}.

\begin{figure}[htbp]
    \centering
    \hspace{0cm}
    \includegraphics[width=7.8cm]{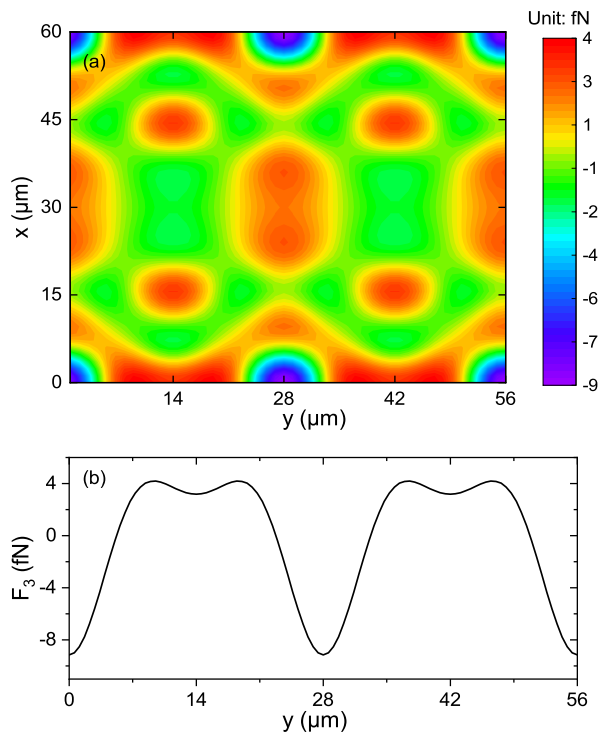}
    \caption{(a) The expected $V_3$ force map in the $xy$-plane at a constant probe-source distance. (b) The $V_3$ force varies along the $y$-axis at $x=0$. In the calculation, $g_p^eg_p^e/4\pi \hbar c$ is set to $1.0\times 10^{-8}$ with $\lambda=10\,\mu$m.}
    \label{fig:v3_force}
\end{figure}

\begin{figure}[htbp]
    \centering
    \includegraphics[width=8.0cm]{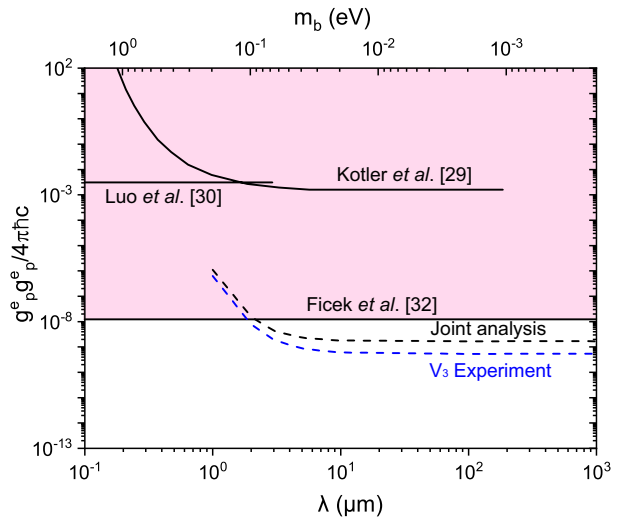}
    \caption{The constraints on the coupling constant of the $V_3$ potential. The blue dashed line represents the lower bound of the parameter space that the proposed experiment can probe assuming only $V_3$ exists. The black dashed line represents the lower bound assuming both the $V_2$ and $V_3$ could exist.}
    \label{fig:v3_coupling}
\end{figure}

The $V_3$ force, depending both on $x$ and $y$, can be calculated numerically. Fig. \ref{fig:v3_force} (a) shows an expected force map for $g_p^eg_p^e/4\pi \hbar c = 1.0\times 10^{-8}$, $\lambda=10\,\mu$m.  As expected, the force is periodic in both the $x$- and $y$-direction. Similar to the $V_2$ search, we plan to modulate the $V_3$ force to the harmonics of the driving frequency by oscillating the source in the $y$-direction. By acquiring data at different points on a plane with a constant probe-source distance, we will obtain a map of force amplitude at the harmonic frequency. The maximum likelihood method can be used to determine the coupling constant for every $\lambda$ by comparing the experimental data with the expected theoretical values, as we have done previously\cite{Wang2016, Ding2020, Ren2021}. Assuming that the experimental results are limited by the minimum detectable force, we can obtain the lower bound of the coupling constant that can be explored in this experiment. As shown in Fig. \ref{fig:v3_coupling}, more than an order of magnitude improvement in $V_3$ detection can be achieved at $\lambda=10\,\mu$m.

If we consider more generally that both the $V_2$ and $V_3$ interactions may exist, we can first determine the coupling constant for the $V_2$ interaction as the stripe source structure is insensitive for the $V_3$ detection. With the $V_2$ coupling constant, we can subtract the $V_2$ interaction in the $V_3$ experiment to get the coupling constant of the $V_3$ interaction. If no signal of new interaction is observed in both experiments, a joint data analysis would yield a limit on the $V_3$ coupling constant, which is approximately 3 times higher than that only one interaction is considered.

\section{Spurious forces}\label{error analysis}
To perform experiments with precision limited by the minimum detectable force of the cantilever, we need to suppress spurious forces to a negligible level. The dominant spurious forces in the experiments are the magnetic force, Casimir force, and electrostatic forces. We will discuss them one by one in the following paragraphs. 

\subsection{Magnetic force}\label{spurious, magnetic force}
\begin{figure}[htbp]
    \centering
    \includegraphics[width=8.0cm]{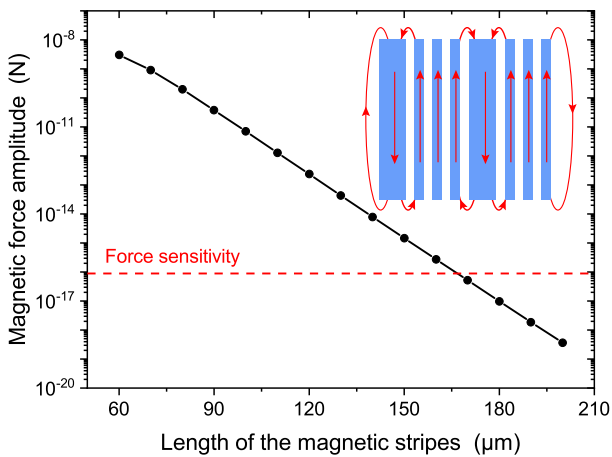}
    \caption{The dependence of the magnetic force amplitude on the length of the magnetic stripes. The minimum detectable force is also presented as the dashed line. Inset: schematic drawing of the magnetic field lines generated by the stripes.}
    \label{fig:v2_mag}
\end{figure}

For the search for spin-spin interactions, the magnetic force between the two objects is the main spurious effect to be considered. In the search of the $V_2$ interaction, we evaluate the magnetic force by numerically integrating the magnetic dipole-dipole interaction between two spins, given by
\begin{equation}
    V_{m}=-\frac{\mu_0 \gamma_e^2 \hbar^2}{16\pi r^3}\left[3\left(\hat \sigma_1 \cdot \hat{r} \right) \left(\hat \sigma_2 \cdot \hat{r} \right)-\left(\hat \sigma_1 \cdot \hat \sigma_2 \right)\right] \label{magnetic force},
\end{equation}
where $\mu_0$ is the vacuum permeability, and $\gamma_e$ is the gyromagnetic ratio of electron. The magnetic force varies periodically with the stripe structure, but its peak-to-peak value decreases rapidly with the length of the stripes,  as shown in Fig. \ref{fig:v2_mag}. The reason is that the magnetic field generated by the stripes is mainly in-plane and closed at the end of stripes (see inset of  Fig. \ref{fig:v2_mag}), thus the magnetic field is negligibly small at the probe's location that is in the center and near the surface of the source structure. The magnetic force is shown to be smaller than the minimum detectable force when the length of the stripe is longer than 170 $\mu$m. Since the real length of the magnetic stripes will be 6 mm, the magnetic force is expected to be much below the minimum detectable force. 

\begin{figure}[htbp]
	\centering
	\includegraphics[width=8.0cm]{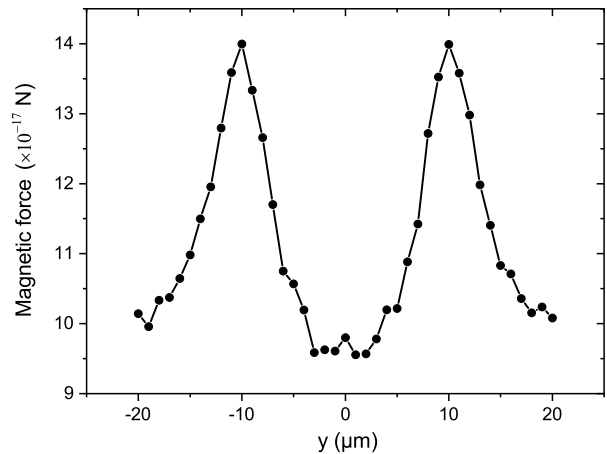}
	\caption{The calculated magnetic force between the magnetized cubes and the superconducting shielded CLMS with the FEA. The volume of the cube is $150\times150\times150$ nm$^{3}$.}
	\label{fig:extrablocks_Fmag}
\end{figure}

The imperfections in the fabrication of the stripes may generate an unexpected magnetic field around the superconducting film-coated CLMS, thus inducing a magnetic force. In order to get a simple idea of how large the force can be, we simulate the imperfections with an array of magnetic cubes. The gaps between the cubes are set to the period of the magnetic stripes in the $y$-direction and the length of the CLMS in the $x$-direction, thus one imperfection exists in the area of a CLMS. Their magnetization is set to 800 kA/m along the $z$-direction to generate maximum magnetic force. We evaluate the force between the magnetic cubes and the superconducting shielded CLMS with the FEA, and find that the volume of the cube should not exceed $\sim150\times150\times150$ nm$^{3}$ to make the force amplitude lower than the minimum detectable force, as shown in Fig. \ref{fig:extrablocks_Fmag}.

\begin{figure*}[hbtp]
    \centering
    \includegraphics[width=17.0cm]{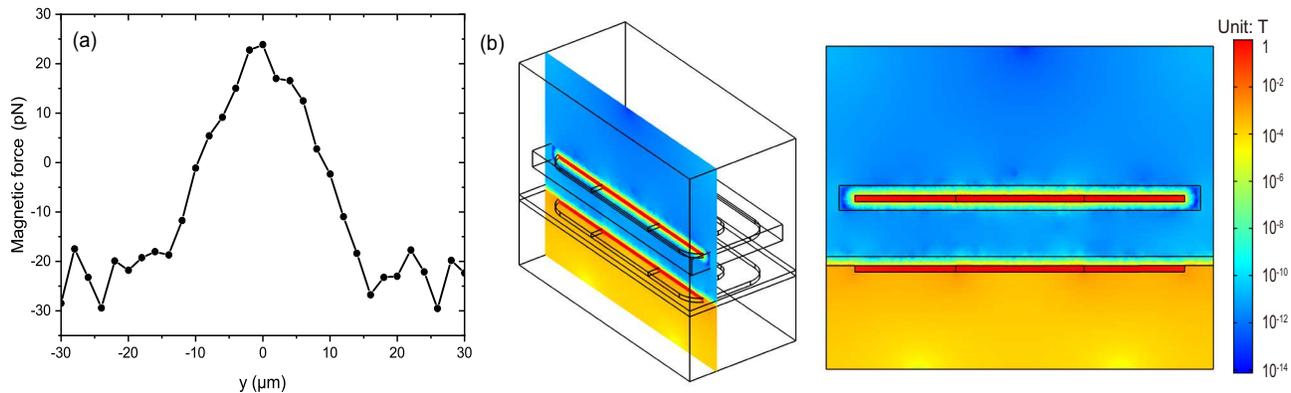}
   \caption{(a) The FEA simulation of the magnetic force between two CLMSs as a function of lateral relative position. (b) Magnetic flux density distribution around the superconducting shielded CLMS on the cantilever and the source.}
    \label{fig:v3_mag}
\end{figure*}

In the search of the $V_3$ interaction, the magnetic force is evaluated with the FEA. We first calculate the lateral position dependence of the magnetic force between two CLMSs at a distance of $d=10\,\mu$m, the result is shown in Fig. \ref{fig:v3_mag} (a). Due to the closed-loop design, the peak magnetic force is reduced to $\sim 10^{-11}$ N, but is still much larger than the minimum detectable force. To further suppress the magnetic force, we propose to encapsulate the CLMSs with superconducting thin films. The closed-loop design reduces the stray field down to the critical field of the superconductor, and then makes superconducting magnetic shielding possible. The magnetic shielding effect is simulated with the FEA (see details in Appendix). According to the simulation, a 1.5-$\mu$m-thick superconducting film can effectively shield the magnetic field down to $2.4\times10^{-11}$T [see  Fig. \ref{fig:v3_mag} (b)]. The magnetic force acting on the cantilever is then reduced to $6.0\times10^{-26}\,$N, which is supposed to be limited by the FEA calculation precision.

Magnetic shielding requires the NbTi film to be superconductive, thus requires the superconducting critical current larger than $1.0\times10^{9}\,$A/$m^{2}$ for a coating thickness of 1.5 $\mu$m according to the FEA simulation. The requirement for the critical current is usually achievable for a NbTi film. If a thicker superconducting film is used, the requirement for the critical current will be less stringent.  On the other hand, we also require that the magnetic field is lower than the lower critical field of the NbTi film at the interface between the magnetic loop and the superconducting thin film. This requires that the permanent magnet film should not be thicker than the soft magnet film to make the magnetic field lower than the lower critical field, which is around 73 mT\cite{Naour1998}.

\subsection{Casimir force}
The Casimir force is mainly contributed from the surface layer of material, where a layer of thickness $d$ contributes about $(1-e^{-4\pi d/\lambda_p}$) of the Casimir force between two infinitely thick metallic plates\cite{Decca2005, Matloob2001}, here $\lambda_p$ is the plasma wavelength of the material. In the proposed experiments, the source structures are either coated with 150-nm-thick gold or 1.5-$\mu$m-thick superconducting thin films. For 150-nm-thick gold film, $e^{-4\pi d/\lambda_p}  \sim 10^{-6}$, which means that the Casimir force difference due to different materials under the coating should be smaller than $10^{-21}$ N.  Thus, here we focus on the variation of the Casimir force due to surface corrugation. The Casimir force is then estimated by proximity force approximation (PFA)\cite{Bordag2009, Blocki1977}. The Casimir energy between two surfaces at a short distance can be approximated as
\begin{equation}
    U^{Ca}=\iint_{D} E_{pp}(z)\mathrm{d}x\mathrm{d}y \label{casimir_fomular},
\end{equation}
where $D$ stands for the projection of the tip to the $xy$-plane with $x$ and $y$ being the integral variables, $E_{pp}(z)$ stands for the Casimir energy per unit area of two electrically neutral, infinitely large, parallel conducting planes at a distance of $z$. Here we use $E_{pp}(z)=-\pi^2\hbar c/720 z^3$, the Casimir energy density of a perfect conductor, for conservative estimation.

To estimate the component of the same period as the source structure, the source surface is modeled as $z=z_0 + z_1$sin$(2\pi y/\Lambda)$, where $z_0$ is the mean level of the surface, $z_1$ is the surface wave amplitude, and $\Lambda$ is the source structure period. In previous experiments, the periodic variation in surface height could be reduced to 3 nm using a SOI-wafer-based fabrication process\cite{Ren2021}. With $z_1 = 3$ nm, we estimate the variation amplitude of the Casimir force between the tip and source structure to be $6.9\times10^{-20}\,$N at a tip-surface distance of $2\, \mu$m. The variation of the Casimir force acting on the source structure is $2.1\times10^{-19}$ N by the surface of the CLMS, and $3.5\times10^{-20}$ N by the rest area of the cantilever. All of the above are much smaller than the minimum detectable force.

\subsection{Electrostatic force}
The electrostatic force is another important spurious force that exists in many precision measurement experiments\cite{Robertson2006, Kim2010Patch, Speake1996, Yin2014}. Similar to the Casimir force, we concern more about the spatially varying force component of the same period as the source structure. These components may arise from the surface corrugation associated with the periodic structure, or from surface patch potential. Since the structures are complicated, here we employ the FEA to calculate the electrostatic force. 

We use the same surface model and tip-surface distance as in the Casimir force calculation. The average residual potential difference can be compensated to around 2 mV by applying a voltage between the tip and the source. The variation amplitude of the electrostatic force is then estimated to be $6.5\times10^{-20}$ N between the tip and the source, $5.3\times10^{-20}$ N between the CLMS on the cantilever and the source, and $7.0\times10^{-19}$ N between the rest area of the cantilever and the source. We see that the variation contributed from the surface corrugation is much smaller than the minimum detectable force.

Patch surface charges are generally randomly distributed over the surface, but their distribution may have the component of the same period as the source structure. To estimate this contribution, we assume that the source surface potential is described as $V(x,y) = V_0 + V_1$sin$(2\pi y/\Lambda)$ referring to the tip. Here $V_0$ is the average potential difference after compensation, and $V_1$ is the potential fluctuation on the source surface. Based on the calculation, in order to make the patch electrostatic force less than the minimum detectable force, we need to make a flat clean surface with a potential fluctuation of less than 1 mV, where the variation amplitude of the electrostatic force is $1.4\times10^{-17}$ N between the tip and the source, $4.0\times10^{-17}$ N between the CLMS on the cantilever and the source, and $1.9\times10^{-17}$ N between the rest area of the cantilever and the source. The actual electrostatic force can be evaluated using data obtained by atomic force microscopy and Kelvin probe force microscopy (KPFM).  The commercially available KPFM can measure the surface potential with a precision of $\sim$ 1 mV and a lateral resolution of $\sim$ 10 nm \cite{Yasutake1995}. Using a gold-coated microsphere as the probe could improve the potential measurement precision,  but still with enough lateral resolution around $\mu$m, which is plausible for the patch electrostatic force evaluation.

\begin{table}
\caption{Experimental parameters used in the proposed experiments.}
  \centering
\begin{tabular*}{\hsize}{@{}@{\extracolsep{\fill}}lccc@{}}
\hline\hline
Parameter                                                               & Value                           & Unit                 \\\hline
CLMS                                                                                                                            \\
$\quad$ length of outer loop                                    & 52    	                	     & $\mu$m               \\
$\quad$ width of outer loop                                     & 20     	                	     & $\mu$m               \\
$\quad$ length of inner loop                               & 40    	                	     & $\mu$m               \\
$\quad$ width of inner loop                                            & 8    	                	     & $\mu$m               \\
$\quad$ thickness                                             & 1    	                	     & $\mu$m               \\
Spin-polarized source in $V_2$ experiment                                                                                                                              \\
$\quad$ length of magnetic stripes                                                       & 6                          & mm               \\
$\quad$ width of wide stripes                                                        & 6                            & $\mu$m               \\
$\quad$ width of narrow stripes                                                       & 2                            & $\mu$m               \\
$\quad$ gap between the stripes                                                       & 2                            & $\mu$m               \\
$\quad$ thickness of stripes                                                    & 1                             & $\mu$m             \\
Spin-polarized source in $V_3$ experiment                                                                                                                              \\
$\quad$ distance between CLMSs ($x$-direction)                                                    & 8                             & $\mu$m               \\
$\quad$ distance between CLMSs ($y$-direction)                                                   & 8                             & $\mu$m               \\
Probe

\\
$\quad$ diagonal length of tip                                                     & 6                             & $\mu$m               \\
$\quad$ tip height                                                    & 10                             & $\mu$m               \\
$\quad$ cantilever length                                                    & 450                            & $\mu$m               \\
$\quad$ cantilever width                                                    & 48                             & $\mu$m               \\
$\quad$ cantilever thickness                                                     & 1                             & $\mu$m               \\
Spin-source distance                                                     & 10                             & $\mu$m               \\
Number density of polarized electrons                                                   & $6.6\times 10^{28}$                             &                \\
\hline\hline
\label{tab:parameter}
\end{tabular*}
\end{table}

\section{Conclusion}\label{conclusion}

In conclusion, we have described the experiments to search for the exotic $V_2$ and $V_3$ interactions by measuring the force between a CLMS and different spin-polarized source structures. Several measures have been taken to suppress the spurious magnetic force, including closed-loop magnetic structure design, superconducting magnetic shielding, and periodic spin source structures. The magnetic force, as well as the Casimir force and electrostatic force, are expected to be lower than the minimum detectable force thanks to those special designs. With the force sensitivity of the cantilever operating at low temperature, the proposed experiments are expected to explore the parameter spaces that are about seven orders of magnitude smaller than the current stringent constraints on $V_2$, and one order magnitude smaller for $V_3$. Furthermore, since the $V_2$ experiment is insensitive to the detection of $V_3$ interaction, we can unequivocally determine the strength of $V_2$ , and then perform a joint analysis to obtain the magnitude of the $V_3$ interaction, assuming they can both exist.
 
\section{Acknowledgments}\label{acknowledgements}
We are indebted to Yiqiu Ma for helpful discussion and suggestion. This work was supported by the National Key R$\&$D Program of China (grant no. 2022YFC2204100)
and the National Natural Science Foundation of China (grants nos. 11875137 and 91736312).

\appendix
 
\section{Simulation of superconducting shielding effect}\label{superconducting shielding effect}
The magnetic shielding effect is simulated with COMSOL Multiphysics. In the superconducting region, we implement the equation combing Ampere’s and Faraday’s laws for the magnetic field $\boldsymbol{H}$\cite{Arsenault2021}, given by
\begin{equation}
   \boldsymbol \nabla  \times \left( {\rho \boldsymbol \nabla  \times \boldsymbol{H} } \right){\rm{ = }} - {\mu _0}\frac{{\partial \boldsymbol{H} }}{{\partial t}},
\end{equation}
where $\rho$ is the resistivity. The superconductor is modeled with a nonlinear resistivity\cite{Rhyner1993}
\begin{equation}
    \rho = \frac{E_c}{J_c}\left(\frac{\left|\boldsymbol{J}  \right|}{J_c}\right)^{n-1},
\end{equation}
where $\boldsymbol{J}$ is the current density, $J_c$ is the critical current density, $n = 40$ is the power law exponent, and $E_c$ = 1 $\mu$V/cm is the critical electrical field. We take $1.0\times10^{10}\,$A/$m^{2}$ as the $J_c$ value, which is usually achievable for NbTi films\cite{Takeda2001}.

For the non-superconducting region, we use the magnetic scalar potential $\phi$ defined as $\boldsymbol{H} =  -\boldsymbol\nabla{\phi}$, the equation to be solved is $\boldsymbol\nabla\cdot\boldsymbol\nabla\phi = 0$. The permanent magnet is modeled with a magnetization of 800 kA/m, and the soft magnet is modeled with a relative magnetic permeability of 8000. A minimum thickness of ~1.5 $\mu$m is determined for the superconducting film to shield the magnetic force. To simulate the periodic structures, we apply periodic boundary conditions in the $x$- and $y$-direction. The magnetic field can be solved by setting appropriate boundary conditions for magnetic field and magnetic flux density. The magnetic force acting on the cantilever is calculated by integrating the Maxwell stress tensor over the outer surface of the superconducting film on the cantilever.


\providecommand{\noopsort}[1]{}\providecommand{\singleletter}[1]{#1}%

\end{document}